\documentclass[10pt]{article}

\usepackage{amsmath}
\usepackage{amsfonts}
\usepackage{amstext}
\usepackage{amssymb}
\usepackage{graphicx}
\usepackage{psfrag}
\usepackage{color}

\begin{document}

%
%  definition
%
\newcommand{\C}{{\mathbb C}}
\newcommand{\qed}{\hfill$\square$}

%\pagestyle{myheadings}
%\markboth{A. Miyake}
%{Multipartite Entanglement under SLOCC}

\title{\Large\bf
Multipartite Entanglement under Stochastic Local Operations
and Classical Communication}

\author{{\large Akimasa Miyake} \medskip\\ 
{\normalsize\it Department of Physics, Graduate School of Science,}
{\normalsize\it University of Tokyo,} \\ 
{\normalsize\it Hongo 7-3-1, Bunkyo-ku, Tokyo 113-0033, Japan} \\
{\normalsize\tt miyake@monet.phys.s.u-tokyo.ac.jp}}

\date{\normalsize 6 January 2004}

%%%%%%%%%%%%%%%%%%%%%%%%%%%%%%%%%%%%%%%%%%%%%%%%%%%%%%%%%%%%
% You may repeat \author \address as often as necessary    %
%%%%%%%%%%%%%%%%%%%%%%%%%%%%%%%%%%%%%%%%%%%%%%%%%%%%%%%%%%%%

\maketitle

%\begin{history}
%\received{(1 November 2003)}
%\revised{(6 January 2004)}
%\end{history}

\begin{abstract}
Stochastic local operations and classical communication (SLOCC),
also called local filtering operations, are a convenient, useful
set of quantum operations in grasping essential properties of
entanglement.
We give a quick overview about the characteristics of multipartite 
entanglement in terms of SLOCC, illustrating the 2-qubit and the rest
($2 \times 2 \times n$) quantum system.
This not only includes celebrated results of 3-qubit pure states, 
but also has implications to 2-qubit mixed states.

\end{abstract}

%\keywords{multipartite entanglement; stochastic LOCC (SLOCC).}

%%%%%%%%%%%%%%%%%%%%%%%%%%%%%%%%%%%%%%%%%%%%%%%%%%%%%%%%%%%%
% The main text of your paper   begins here                %
%%%%%%%%%%%%%%%%%%%%%%%%%%%%%%%%%%%%%%%%%%%%%%%%%%%%%%%%%%%%

\section{Introduction}

Quantum information science would offer us remarkably new
information society.
There, quantum correlation called entanglement enables us 
to process informational tasks which are less efficient or
impossible in the classical manner.
Entanglement is expected to be especially intriguing and valuable 
in the multi-party setting, since the network nature (i.e., interactions
of many elements) is essence of information society.

Entanglement is quantum correlation, which is never increasing
on average by local operations and classical communication (LOCC)
\cite{bennett96}. 
In other words, LOCC will preserve or destroy entanglement. 
For the {\it bipartite} pure states, these monotone properties
of entanglement under LOCC have been successfully characterized 
both in the single copy case and in the asymptotic (infinitely many, 
identical copies) case.
In particular, for the single copy situation, the LOCC equivalence class of 
entanglement turns out to be reduced to the equivalence class under 
local unitary operations (LU), in other words, the equivalence class of 
states with same coefficients in the Schmidt decomposition
(the normal form in terms of a biorthonormal product basis) 
\cite{vidal00,bennett00}.
Then, monotone properties of different entangled states
under LOCC are characterized by a majorization principle of 
the Schmidt coefficients \cite{lo01,nielsen99}.

However, the studies of multipartite entanglement have been found
very challenging.
Since the useful techniques in the bipartite situation, such as the Schmidt 
decomposition, cannot be generalized to the multipartite situation 
straightforward, it is difficult to capture even the LOCC equivalence
class, and certainly the monotonicity of entanglement.
Alternatively, the LU equivalence class of entanglement, which is by 
definition finer than the LOCC equivalence class, has been studied
for the 3-qubit pure states \cite{linden98_3bit}.
Still, the convertibility (monotonicity) between these LU equivalence
classes has not yet been unveiled.
The situation partly motivates the introduction of our SLOCC equivalence.
As another motivation, it should be noted that we are very interested 
in the genuine "multipartite" phenomena (such as the limited shareability 
of multipartite entanglement), which cannot be fully captured by dividing 
multiparties into bipartite subgroups.  

Here, we will address the stochastic LOCC (SLOCC) \cite{bennett00,dur00}
classification of entanglement, which is 
a coarse-grained classification of entanglement under LOCC. 
{\it An advantage of the SLOCC classification lies in the fact that we can
grasp the essential nature of entanglement, by successfully characterizing 
the equivalence class (conservation) of entanglement and monotone 
(irreversible) properties of entanglement under LOCC.}

\section{Stochastic LOCC}

Let us consider the single copy of a
multipartite pure state $|\Psi\rangle$ on the finite dimensional,
$l$-partite Hilbert space
${\mathcal H} = \C^{k_1}\otimes \cdots \otimes \C^{k_l}$
(precisely, we consider a ray on its complex projective 
Hilbert space $\mathcal{P} (\C^{k_{1}} \otimes\cdots\otimes \C^{k_{l}})$),
\begin{equation}
\label{eq:psi}
|\Psi\rangle =\sum_{i_1,\ldots,i_l =0}^{k_1 -1,\ldots,k_l-1} 
\psi_{i_1 \ldots i_l} |i_1\rangle \otimes\cdots\otimes |i_l\rangle,
\end{equation}
where a set of $|i_1\rangle \otimes\cdots\otimes |i_l\rangle$
constitutes the standard computational basis and is often
abbreviated to $|i_1 \cdots i_l\rangle$. 
In SLOCC, we identify two states $|\Psi\rangle$ and $|\Psi'\rangle$,
which are interconvertible with nonvanishing probabilities 
(the success probability of the conversion may differ in back and forward
directions), as equivalent entangled states.
On the other hand in LOCC, we ordinarily identify two states interconvertible
deterministically (with probability 1) as equivalent ones,
as seen in the Section~1.
Mathematically, two SLOCC equivalent entangled states $|\Psi\rangle$ 
and $|\Psi'\rangle$ are connected by {\it invertible} local operations 
\cite{dur00},
\begin{equation}
\label{eq:slocc}
|\Psi'\rangle = M_1 \otimes \cdots \otimes M_l \; |\Psi\rangle, \qquad
|\Psi\rangle  = M_1^{-1} \otimes \cdots \otimes M_l^{-1} \; |\Psi'\rangle, 
\end{equation}
where $M_i$ is any local operation having a nonzero determinant
on the $i$-th party, i.e., $M_i$ is an element of the general 
linear group $GL_{k_i}(\C)$ (we do not care about
the overall normalization and phase so that we can take its
determinant 1, i.e., $M_i \in SL_{k_i}(\C)$). 
The point is a nice group structure of SLOCC; i.e., SLOCC
would practically consist of successive rounds of measurements and 
communication of their outcomes, but the whole is given simply
by one "large" local operation of Eq.~(\ref{eq:slocc}).
It can be also said that SLOCC are a set of trace decreasing, 
completely positive maps by the postselection of a sequence of 
successful outcomes.
In short, the SLOCC classification of multipartite entanglement 
means clarifying the moduli space under the SLOCC equivalence, i.e.,
the structure of orbits generated by a direct product of special
linear groups $SL_{k_1}(\C) \times\cdots\times SL_{k_l}(\C)$. 

Afterwards, we will address the question of {\it noninvertible} local 
operations (at least one of the ranks of $M_i$ in Eq.~(\ref{eq:slocc}) is
not full, so that no inverse operations exist), in order to characterize 
the order between different SLOCC equivalence classes. 
By definition, the probabilistic conversion between them would be possible 
only in one direction, or would be impossible in both directions.
The set of invertible and noninvertible local operations are often called 
local filtering operations \cite{gisin96,linden98}.

\begin{figure}[t]
\begin{center}
\includegraphics[width=8cm,clip]{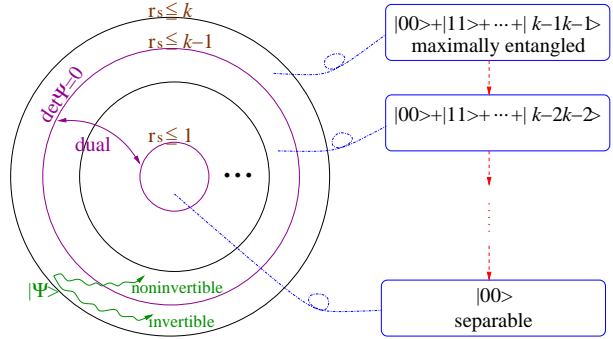}
\caption{(Left) The onion-like classification of SLOCC orbits
in the two $k$-level ($k \times k$) quantum system.
The onion skins represent the orbit closures, and the orbits (between
the onion skins) are different entangled classes.
Note that the largest subset $\overline{{\mathcal O}_{k-1}}$ of 
non generic entangled states is dual to the smallest subset 
$\overline{{\mathcal O}_{1}}$ of separable states.
(Right) The structure of bipartite entangled pure states under 
noninvertible local operations.
SLOCC entangled classes are totally ordered in such a way that 
an entangled class of the larger rank is more entangled than that
of the smaller one.}
\label{fig:bipartite}
\end{center}
\end{figure}

In the bipartite (for simplicity, $k_1 = k_2 =k$) case, the SLOCC 
classification means just classifying the whole states
by the so-called Schmidt local rank (equivalently,
the rank of the coefficient matrix $\Psi = (\psi_{i_1 i_2})$
of Eq.~(\ref{eq:psi})).
We readily see that the rank is the SLOCC invariant,
because $\Psi$ is transformed as
\begin{equation}
\Psi' =  M_1 \Psi  M_2^T,
\end{equation}
under an invertible local operation 
$G = M_1 \otimes M_2 \in SL_k\times SL_k$ 
(the superscript $T$ stands for the transposition).
Although, in this bipartite case, the observation immediately suggests that
there exist $k$ different entangled classes distinguished by 
SLOCC-invariant rank $r_s$, let us put the problem on the following picture,
for later convenience.
A set $\overline{{\mathcal O}_{r_s}}$ of states of the local rank 
less than or equal to $r_s$ is a {\it closed} subset under SLOCC, and 
$\overline{{\mathcal O}_{{r_s}-1}}$ is the singular locus of 
$\overline{{\mathcal O}_{r_s}}$. 
This is how the local rank leads to an "onion" structure  
(mathematically the stratification) of Fig.~\ref{fig:bipartite}:
\begin{equation}
\label{eq:onion}
\overline{{\mathcal O}_{k}} \supset 
\overline{{\mathcal O}_{k-1}} \supset\cdots\supset 
\overline{{\mathcal O}_1} \supset 
\overline{{\mathcal O}_0} = \emptyset, 
\end{equation}
and the orbits ${\mathcal O}_{r_s}= \overline{{\mathcal O}_{r_s}} - 
\overline{{\mathcal O}_{{r_s}-1}} \;(r_s=k,\ldots,1)$ give $k$ classes of 
entangled states.

Now we discuss the relationship between these classes under 
noninvertible local operations.
Since the local rank only decreases by noninvertible local operations 
\cite{lo01}, we find that $k$ classes are totally ordered in such a way 
that an entangled class of the larger rank is more entangled
than that of the smaller rank.
In particular, 
the outermost generic set ${\mathcal O}_{k}$ can be called the class of 
maximally entangled states, since this class can convert to all classes 
by LOCC but other classes never convert to it.
On the other hand, the innermost set ${\mathcal O}_{1}$ is indeed the class 
of separable (non-entangled) states, since this class never convert to 
other classes by LOCC but any other classes can convert to it.
Note that, in the multipartite situation, a set of the SLOCC-invariant
local ranks, as a straightforward generalization of the Schmidt local rank, 
is not sufficient to classify entangled classes completely.
Still, the onion structure of SLOCC orbit closures, as described above, 
will be indeed utilized in the generalization to the multipartite situation.

\section{Characteristics of multipartite entanglement}

Let us start with accounting for why we are interested in
multipartite entanglement in the 2-qubit and the rest
($2 \times 2 \times n$) quantum system here.

1. This case includes 4-qubit entanglement distributed over 
{\it 3 parties}, for example. We readily see that two 
Einstein-Podolsky-Rosen (EPR)-Bell pairs over 3 parties can create 
the so-called Greenberger-Horne-Zeilinger (GHZ) 3-qubit entangled state,
and in turn the GHZ state can create one Bell pair under LOCC.
Observing that these multi-party LOCC protocols suggest the transformation
between bipartite entanglement and tripartite entanglement,
we can see that our situation gives a minimal setting for multi-party LOCC 
protocols.
Then, we would ask, what kinds of essentially different 
multipartite entanglement there exist in this situation?

2. Our case has implications to 2-qubit mixed states.

3. As a bit mathematical remark, although there are 
infinitely many orbits in general even under the stochastic LOCC 
\cite{dur00,miyake03},
our case will turn out to be finite so that it is amenable to treat.
The reason is as follows; the dimension of the Hilbert space expands
exponentially as the number of parties $l$ increases, while
the dimension of the local operations we can perform increases just linearly.
Precisely, the number of nonlocal complex parameters remained in the 
generic position of the projective Hilbert space (the space of rays)
is given by
\begin{equation}
\label{eq:nonlocal_para}
{\rm dim} \frac{\mathcal{P}(\C^{k_1}\otimes\cdots\otimes\C^{k_l})}
{SL_{k_1}\times\cdots\times SL_{k_l}} 
= (k_1 \times\cdots\times k_l -1) - ((k_1^2 -1)+ \cdots + (k_l^2 -1) 
- \delta),
\end{equation}
where $\delta$ is the dimension of the stabilizer of the SLOCC
operations $SL_{k_1}\times\cdots\times SL_{k_l}$ there.
So, if Eq.~(\ref{eq:nonlocal_para}) is strictly positive then
there are infinitely many orbits in the generic (maximal dimensional) 
class, otherwise there are just finitely many orbits.
For example, in the 4-qubit ($2 \times 2 \times 2 \times 2$) case,
Eq.~(\ref{eq:nonlocal_para}) is $15 - 12 = 3 > 0$ so that 3 nonlocal
parameters remain in the generic class \cite{verstraete02_4bit,miyake03}.
In contrast, in the $2 \times 2 \times 4$ case, 
Eq.~(\ref{eq:nonlocal_para}) is $0$ and this case consists of 
finite orbits.

We utilize two, algebraic and geometric ideas to achieve our goal.
Although we illustrate $2 \times 2 \times n$ cases here (see also 
Ref.~\cite{mi_ver03} in detail), it can be seen that these ideas 
are useful to other cases.

1. Observing an homomorphism $SL_2(\C)\otimes SL_2(\C)\simeq SO_4 (\C)$, 
and concatenating the indices of Alice and Bob
as the "flattened" (2-dimensional) matrix $\tilde{\Psi}=(\psi_{(i_1 i_2)i_3})$,
let us consider $R$, defined as
\begin{equation} 
\label{eq:R}
R = T \, \tilde{\Psi},  \quad 
T=\frac{1}{\sqrt{2}}\left(\begin{array}{cccc}
1&0&0&1\\ 0&i&i&0\\ 0&-1&1&0\\ i&0&0&-i
\end{array}\right),
\end{equation}
in stead of the "3-dimensional" matrix  
$\Psi=(\psi_{i_1 i_2 i_3})$ in Eq.~(\ref{eq:psi}).
According to Eq.~(\ref{eq:slocc}), our problem is found to be 
algebraically equivalent to looking for normal forms of the 
"ordinary" matrix $R$ under left multiplication of an element 
$O$ in $SO_4 (\C)$ and right multiplication of an element 
$M_3^T$ in $SL_n (\C)$, i.e.,
\begin{equation}
R' = O \, R \, M^{T}_{3}. 
\end{equation}

2. Geometrically speaking, we focus on a duality, a generalization 
of the Legendre transformation, between the original state space and
its conjugate state space.
Associated with the duality between the set of separable states 
and its dual set of degenerate entangled states,
hyperdeterminants appear as key SLOCC invariants in the classification
of multipartite entanglement (see Ref.~\cite{miyake03} in detail).
Indeed, in the bipartite ($k_1 = k_2 =k$) case (cf.  
Fig.~\ref{fig:bipartite}), the set of separable states
is given by the set of states of the rank $1$, i.e., 
by $\overline{{\mathcal O}_1}$.
On the other hand, its dual set is given by the set of all degenerate 
entangled states where their rank is not full ($\det \Psi =0$), i.e., 
by $\overline{{\mathcal O}_{k-1}}$. 
Remarkably, the duality also holds for the multipartite case, and 
the dual set is given, in analogy, by the zero hyperdeterminant: 
${\rm Det}\Psi =0$. 
Note that  the hyperdeterminant for the 3-qubit system has been also known 
as the 3-tangle \cite{coffman00}, and represents the genuine tripartite
component in the 3-qubit entanglement sharing (cf. Appendix~A).
Generally, the absolute value of a hyperdeterminant is the entanglement 
measure (precisely, entanglement monotone) which represents the amount of 
generic (non-degenerate) entanglement for the given format (cf. Section~4).

\begin{figure}[t]
\begin{center}
\includegraphics[width=5cm,clip]{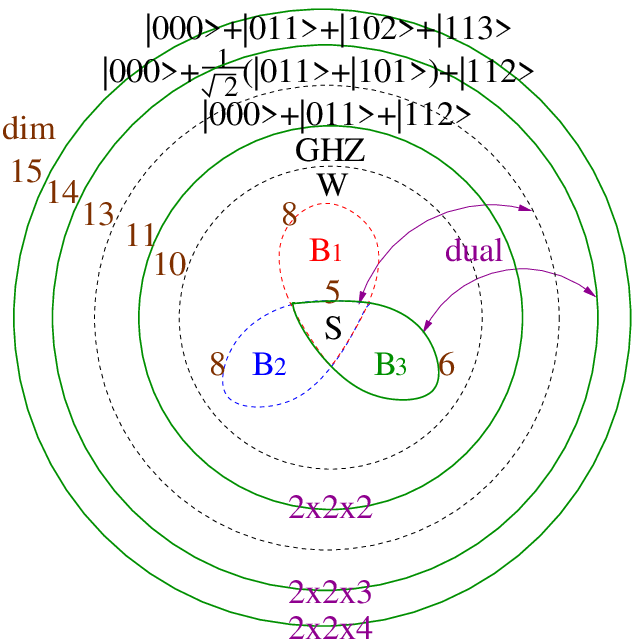}  % 5.5
\hspace{0.5cm}
\includegraphics[width=6cm,clip]{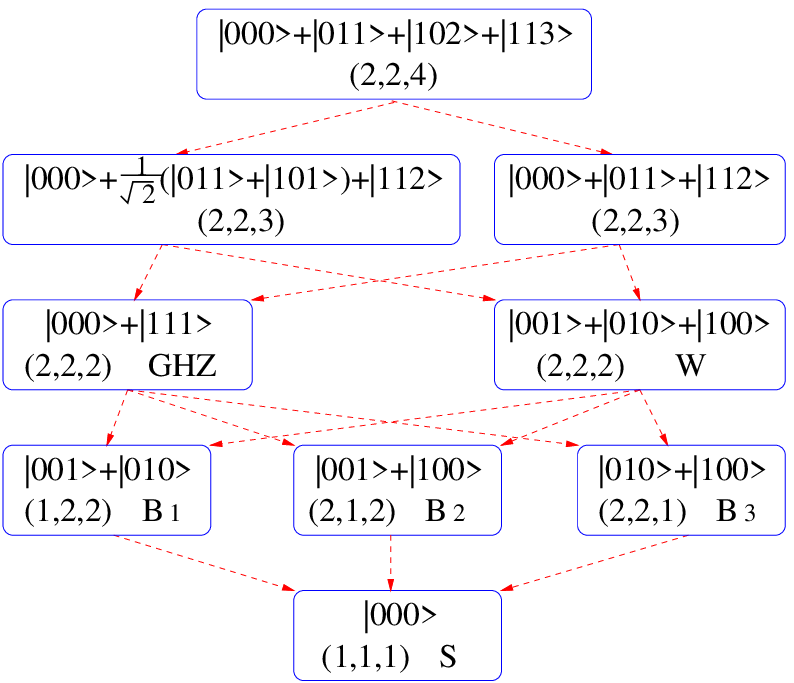}      % 6
\caption{
(Left) The onion-like classification of SLOCC orbits 
in the 2-qubit and the rest ($2 \times 2 \times n $ ($n\geq 4$)) 
quantum system.
There are nine different SLOCC entangled classes.
(Right) The five-graded partially ordered structure of nine entangled classes.
Every class is labeled by its representative, its set of local ranks, and
its name. Noninvertible local operations, indicated by dashed arrows, degrade
higher entangled classes into lower entangled ones.
Both figures partly include the cases for $n=3$ and $2$.}
\label{fig:part}
\end{center}
\end{figure}

Using these ideas, we obtain the classification of 
$2 \times 2 \times n$ multipartite entanglement, drawn in 
Fig.~\ref{fig:part}.
For $n \geq 4$, the whole space is divided into nine entangled 
classes (orbits) by {\it invertible} local operations, while for $n=3$ 
or $2$, it is divided into eight \cite{miyake03} or six \cite{dur00} 
classes, respectively.
These essentially different entangled classes constitute a five-graded
partial order under {\it noninvertible} local operations for $n \geq4$,
while a four or three-graded one for $n=3$ or $2$, respectively. 
So, as the third party Clare's subsystem becomes large, the number of 
SLOCC entangled classes increases and the partial order gets higher.
Representative states of the GHZ and W classes
are famous in their different physical properties and applications
to quantum information processing (QIP) \cite{dur00}.
The GHZ state $|000\rangle +|111\rangle$ has the maximal amount of 
generic 3-qubit entanglement measured by 
the hyperdeterminant of the format $2 \times 2 \times 2$ 
(cf. Eq.(\ref{eq:hdet222}) and Appendix~A).
It violates the Bell's inequality maximally, and enable us to extract 
one Bell pair between any two parties out of three with probability 1.
On the other hand, the W state $|001\rangle+|010\rangle+|100\rangle$
has the maximal amount of average pairwise entanglement distributed
over 3 parties (cf. Appendix~A). 
So, the states in the W class can be utilized in optimal quantum cloning 
\cite{buzek96}.

Remarkably, in the $2 \times 2 \times n \;(n \geq 4)$ cases,
the generic (maximal dimensional) class is a unique
"maximally entangled" class, lying on the top of the partial order.   
So this class serves as an almighty resource that can create any state
probabilistically. The situation is in contrast with the 3-qubit 
($n=2$) case, where there are two inequivalent entangled classes on 
the top of the hierarchy.
In addition, among the generic class, its representative state
(two Bell pairs distributed over 3 parties),
\begin{align}
\label{eq:2bell}
|\mbox{2 Bell}\rangle_{ABC_{12}} &=
\frac{1}{\sqrt{2}}(|00\rangle + |11\rangle)_{A C_{1}} \otimes 
\frac{1}{\sqrt{2}}(|00\rangle + |11\rangle)_{B C_{2}} \nonumber\\
&= \frac{1}{2} (|00(00)\rangle+|01(01)\rangle+|10(10)\rangle+
|11(11)\rangle)_{ABC_{12}},
\end{align}
is most powerful in the sense that it can create any state {\it with 
probability 1} by the Clare's positive operator valued measure (POVM) 
followed by recovery unitary operations of Alice and Bob.
Generally, we find that when one of multiparties holds at least a half 
of the total Hilbert space, the representative state of the generic class 
can create any state with certainty.
So, the situation is somehow analogous to the bipartite cases.

We briefly discuss an interpretation of Fig.~\ref{fig:part} in QIP.
Let us observe that the initially prepared state in entanglement swapping 
\cite{zukowski93} is two Bell pairs (of Eq.~(\ref{eq:2bell})) over 3 parties, 
included in the generic class.
Clare's local collective measurement by the Bell basis 
$|\Phi_\mu\rangle_{C_{12}} \;(\mu=1, \ldots, 4)$ yields with probability
$1/4$ a Bell pair $|\Phi_\mu\rangle_{AB}$, which depends on her 
outcome $\mu$, due to $|{\rm 2 Bell}\rangle_{ABC_{12}} = 
\frac{1}{2}\sum_{\mu=1}^{4} |\Phi_\mu\rangle_{AB} \otimes 
|\Phi_\mu\rangle_{C_{12}}$,
and any Bell pair $|\Phi_\mu\rangle_{AB}$ created between Alice and Bob is 
LOCC equivalent to $\frac{1}{\sqrt{2}}(|00\rangle +|11\rangle)_{AB}$.
Thus, we can say that the entanglement swapping is 
a LOCC protocol creating the isolated (maximal) bipartite entanglement between
Alice and Bob from generic entanglement. 
In other words, entanglement swapping is given by a downward (noninvertible) 
flow in Fig.~\ref{fig:part} from the generic class to the biseparable class 
$B_3$.
We readily see that any state containing tripartite entanglement 
(i.e., all its local ranks are greater than 1) is powerful enough to create 
probabilistically one Bell pair in one of the biseparable classes 
$B_1$, $B_2$, or $B_3$.
On the other hand, two Bell pairs can create two different kinds of the 
3-qubit entanglement, GHZ and W.
These LOCC protocols are given by downward flows from the generic class to
the GHZ and W class, respectively.
That is how we see that important multi-party LOCC protocols in QIP are given
as downward flows in the partially ordered structure, such as
Fig.~\ref{fig:part}, of multipartite entangled classes.
So, the development of SLOCC classification of multipartite entanglement 
can be expected to give us an insight in looking for new LOCC protocols.

\section{Measures for multipartite entanglement}

We give a convenient criterion to distinguish these nine SLOCC classes.
A set $(r_1,r_2,r_3)$ of the local ranks of the reduced density matrices,
defined by 
\begin{equation}
r_i = {\rm rk} \;\rho_i, \qquad 
\rho_i ={\rm tr}_{\forall j\ne i}(|\Psi\rangle\langle\Psi|), 
\end{equation}
is a straightforward generalization of the Schmidt local rank in the bipartite 
case, and indeed is a good set of SLOCC invariants in the 
multipartite case, too.
In the bipartite case, this local rank is necessary and sufficient to 
distinguish SLOCC classes, however in the multipartite case, we need 
additional SLOCC invariants in order to complete the classification.
As we can see in the Table~1, the cases where the local ranks are
$(2,2,3)$ or $(2,2,2)$ are not classified yet.

An option as the additional SLOCC invariants is the rank of $R^T R$
of Eq.~(\ref{eq:R}), associated with the 2-qubit substructure, i.e.,
with the homomorphism $SL_2(\C)\otimes SL_2(\C) \simeq SO_4(\C)$. 
We will discuss this invariant later in the Section~5.
Alternative option is the series of hyperdeterminants adjusted to the smaller
formats. 
These invariants reflect the nature of the onion structure, where the smaller 
format is embedded in the larger format.
In order to classify the $(2,2,3)$ case further, we can utilize the
hyperdeterminant of the format $2 \times 2 \times 3$,
\begin{align}
\label{eq:hdet223}
{\rm Det}_{223}\Psi =&
\left|\begin{array}{ccc}
\psi_{000} & \psi_{001} & \psi_{002} \\ \psi_{010} & \psi_{011} & \psi_{012} \\
\psi_{100} & \psi_{101} & \psi_{102} \\
\end{array}\right|
\left|\begin{array}{ccc}
\psi_{010} & \psi_{011} & \psi_{012} \\ \psi_{100} & \psi_{101} & \psi_{102} \\
\psi_{110} & \psi_{111} & \psi_{112} \\
\end{array}\right| \nonumber\\
& \quad - 
\left|\begin{array}{ccc}
\psi_{000} & \psi_{001} & \psi_{002} \\ \psi_{010} & \psi_{011} & \psi_{012} \\
\psi_{110} & \psi_{111} & \psi_{112} \\
\end{array}\right|
\left|\begin{array}{ccc}
\psi_{000} & \psi_{001} & \psi_{002} \\ \psi_{100} & \psi_{101} & \psi_{102} \\
\psi_{110} & \psi_{111} & \psi_{112} \\
\end{array}\right|,
\end{align}
after adjusting $\psi_{i_1 i_2 i_3}$ to the $2 \times 2 \times 3$ format
(i.e., choosing $\forall \psi_{i_1 i_2 i_3}=0 \;(i_3 \geq 3)$)
by Clare's local operation.
Likewise, in order to classify the $(2,2,2)$ case, we can utilize
the hyperdeterminant of the format $2 \times 2 \times 2$, also called
the 3-tangle \cite{coffman00},
\begin{align}
\label{eq:hdet222}
{\rm Det}_{222}\Psi  
=& \mbox{\;} \psi_{000}^2 \psi_{111}^2 + \psi_{001}^2 \psi_{110}^2
+ \psi_{010}^2 \psi_{101}^2 + \psi_{100}^2 \psi_{011}^2 \nonumber\\ 
& \mbox{} - 2(\psi_{000}\psi_{001}\psi_{110}\psi_{111}+  
 \psi_{000}\psi_{010}\psi_{101}\psi_{111}
 + \psi_{000}\psi_{100}\psi_{011}\psi_{111} \nonumber\\
& \mbox{} +\psi_{001}\psi_{010}\psi_{101}\psi_{110} 
 + \psi_{001}\psi_{100}\psi_{011}\psi_{110}+  
 \psi_{010}\psi_{100}\psi_{011}\psi_{101})  \nonumber\\
& \mbox{} + 4 (\psi_{000}\psi_{011}\psi_{101}\psi_{110} +
 \psi_{001}\psi_{010}\psi_{100}\psi_{111}). 
\end{align}
This polynomial invariant distinguishes the GHZ class and the W class.
Since these hyperdeterminants represent the genuine component
in multipartite entanglement of the given formats, they are also expected 
to play key roles in entanglement sharing in the similar manner as 
the 3-qubit case
(For a short explanation of entanglement sharing, see the Appendix~A).

\begin{table}[t]
%\tbl{Entanglement measures distinguishing SLOCC entangled classes}
{\begin{tabular}{|c|c|c|c|c|}
\hline
SLOCC class & $(r_1,r_2,r_3)$ & $r(R^T R)$ 
& ${\rm Det}_{223}\Psi$ & ${\rm Det}_{222}\Psi$ \\
\hline
$|000\rangle+|011\rangle+|102\rangle+|113\rangle$ &
(2,2,4) & 4 & --------- & --------- \\
\cline{1-4}
$|000\rangle+\frac{1}{\sqrt{2}}(|011\rangle+|101\rangle)+|112\rangle$ &
(2,2,3) & 3 & $\ne 0$ & --------- \\
\cline{1-1}\cline{3-4}
$|000\rangle+|011\rangle+|112\rangle$ & (2,2,3) & 2 & 0 & --------- \\
\cline{1-2}\cline{5-5}
$|000\rangle+|111\rangle$ & (2,2,2) & 2 & 0 & $\ne 0$ \\
\cline{1-1}\cline{3-3}\cline{5-5}
$|001\rangle+|010\rangle+|100\rangle$ & (2,2,2) & 1 & 0 & 0 \\
\cline{1-2}
$|010\rangle+|100\rangle$ & (2,2,1) & 1 & 0 & 0 \\
\cline{1-3}
$|001\rangle+|100\rangle$ & (2,1,2) & 0 & 0 & 0 \\
\cline{1-2}
$|001\rangle+|010\rangle$ & (1,2,2) & 0 & 0 & 0 \\
\cline{1-2}
$|000\rangle$ & (1,1,1) & 0 & 0 & 0 \\
\hline
\end{tabular}}
\caption{Entanglement measures distinguishing SLOCC entangled classes}
\end{table}

Due to their importance as measures of entanglement, 
we show that the absolute value of the hyperdeterminant of format
$k_1 \times \cdots \times k_l$ becomes an entanglement monotone, i.e., 
is decreasing on average under LOCC \cite{vedral97,vidal00}.
This nice property is essentially due to the arithmetic mean-geometric mean 
inequality for (hyper)determinants, as outlined in Refs.~\cite{miyake03}.
The proof is similar to the argument of Ref.~\cite{dur00}, where
the 3-tangle (hyperdeterminant of format $2\times 2\times 2$) was
proved to be an entanglement monotone.

Any LOCC protocol is decomposed into POVMs, each of which is unilocal 
operations of just one party.
Moreover, we restrict ourselves to treat two-outcome POVMs, since
any (local) POVM can be implemented as a sequence of two-outcome POVMs. 
Along with the permutation invariance, up to sign, of hyperdeterminants 
over parties, we assume, without loss of generality, that the
two-outcome POVM is performed by Alice.
Let $M(1)$ and $M(2)$ be the Alice's two POVM elements, satisfying
\begin{equation}
\label{eq:povm_em}
M^{\dag}(1)M(1)+M^{\dag}(2)M(2)=\mathbf{1}.
\end{equation}
By using the singular value decomposition for a $k_1 \times k_1$ matrix 
$M(\mu) \;(\mu=1,2)$, we can write
\begin{align}
%\left\{\mbox{}
%\begin{aligned}
M(1) &= U(1) D(1) V = 
U(1)\left(\begin{array}{ccc} 
\alpha_0 & & O\\ & \ddots & \\ O & & \alpha_{k_1 -1}
\end{array}\right) V , \\
M(2) &= U(2) D(2) V =
U(2)\left(\begin{array}{ccc} 
\beta_0 & & O\\ & \ddots & \\ O & & \beta_{k_1 -1}
\end{array}\right) V , 
%\end{aligned}
%\right.
\end{align}
where $U(\mu) \;(\mu=1,2)$ and $V$ are unitary matrices, and 
$D(\mu) \;(\mu=1,2)$ are diagonal matrices with 
$0 \leq \alpha_i \leq 1$ and $0 \leq \beta_i \leq 1$ satisfying 
$\alpha_i^2 + \beta_i^2 =1$ for $i=0,\ldots,k_1 -1$.
Note that according to Eq.~(\ref{eq:povm_em}), we can choose $V$ to be same
for $M(1)$ and $M(2)$. 

Let an initial pure state before Alice's POVM be $|\Psi\rangle$
with the hyperdeterminant ${\rm Det}\Psi$.
After the POVM, we obtain two outcome (normalized) states 
$|\Psi'(\mu)\rangle \;(\mu=1,2)$ for $M(\mu)$ with the probability $p_\mu$.
In terms of 
\begin{equation}
V|\Psi\rangle = \left(\begin{array}{c} 
z_0 \\ \vdots \\ z_{k_1-1}
\end{array}\right),
\end{equation}
with the normalization condition $|z_0|^2 + \cdots + |z_{k_1 -1}|^2 =1$, 
these outcome states are written as
\begin{align}
%\left\{\mbox{}
%\begin{aligned}
|\Psi'(1)\rangle &= \frac{1}{\sqrt{p_1}} M(1)|\Psi\rangle
=\frac{1}{\sqrt{p_1}} U(1) 
\left(\begin{array}{c}
\alpha_0 z_0 \\ \vdots \\ \alpha_{k_1 -1} z_{k_1 -1}
\end{array}\right), \\
|\Psi'(2)\rangle &= \frac{1}{\sqrt{p_2}} M(2)|\Psi\rangle
=\frac{1}{\sqrt{p_2}} U(2) 
\left(\begin{array}{c}
\beta_0 z_0 \\ \vdots \\ \beta_{k_1 -1} z_{k_1 -1}
\end{array}\right), 
%\end{aligned}
%\right.
\end{align}
where the outcoming probabilities are
\begin{equation}
%\left\{\mbox{}
%\begin{aligned}
p_1 = \alpha^2_0 |z_0|^2 + \cdots + \alpha^2_{k_1 -1} |z_{k_1 -1}|^2 , 
\quad
p_2 = \beta^2_0 |z_0|^2 + \cdots + \beta^2_{k_1 -1} |z_{k_1 -1}|^2 ,
%\end{aligned}
%\right.
\end{equation}
and satisfies $p_1 + p_2=1$. 
Now, we show that the the following inequality, i.e., the defining property
of entanglement monotones,
\begin{equation}
\label{eq:hdet_em}
|{\rm Det}\Psi| \geq p_1 |{\rm Det}\Psi'(1)| + p_2 |{\rm Det}\Psi'(2)|,
\end{equation}
holds for all possible POVM elements $\{M(1),M(2)\}$.
Suppose the degree of homogeneity of the hyperdeterminant is $d$ 
(say, $d=6$ in Eq.~(\ref{eq:hdet223}) and $d=4$ in Eq.~(\ref{eq:hdet222})).
Since the hyperdeterminant is $SL_{k_1}\times \cdots \times SL_{k_l}$ 
invariant, also immediately $SU_{k_1}\times\cdots\times SU_{k_l}$ invariant,
for the normalized pure states, we have
\begin{align}
%\left\{\mbox{}
%\begin{aligned}
{\rm Det}\Psi'(1) &= 
{\rm Det} \frac{1}{\sqrt{p_1}}\left(\begin{array}{c}
\alpha_0 z_0 \\ \vdots \\ \alpha_{k_1 -1} z_{k_1 -1}
\end{array}\right) 
=\frac{\alpha^{d/{k_1}}_0 \cdots 
\alpha^{d/{k_1}}_{k_1 -1}}{\sqrt{p_1}^d}
{\rm Det}\Psi , \\
{\rm Det}\Psi'(2) &= 
{\rm Det} \frac{1}{\sqrt{p_2}}\left(\begin{array}{c}
\beta_0 z_0 \\ \vdots \\ \beta_{k_1 -1} z_{k_1 -1}
\end{array}\right)
=\frac{\beta^{d/{k_1}}_0 \cdots 
\beta^{d/{k_1}}_{k_1 -1}}{\sqrt{p_2}^d}
{\rm Det}\Psi .
%\end{aligned}
%\right.
\end{align}
According to Eq.~(\ref{eq:hdet_em}), all we should show is that
\begin{equation}
\label{eq:hdet_em2}
1 \geq
\frac{\alpha_0^{d/{k_1}} \cdots \alpha_{k_1 -1}^{d/{k_1}}}{\left(
\alpha_0^2 |z_0|^2 + \cdots + \alpha_{k_1 -1}^2 |z_{k_1 -1}|^2
\right)^{\frac{d-2}{2} }} +
\frac{\beta_0^{d/{k_1}} \cdots \beta_{k_1 -1}^{d/{k_1}}}{\left(
\beta_0^2 |z_0|^2 + \cdots + \beta_{k_1 -1}^2 |z_{k_1 -1}|^2
\right)^{\frac{d-2}{2} }} 
\end{equation}
is fulfilled for any set of $\alpha_i$ and $\beta_i =\sqrt{1-\alpha_i^2}$ 
$(i=0,\ldots,k_1-1)$.
Remember the arithmetic mean-geometric mean inequality:
for $q_1,\ldots,q_k \geq 0$, 
\begin{equation}
\frac{q_1 +\cdots + q_k}{k} \geq \sqrt[k]{q_1 \cdots q_k},
\end{equation}
with the equality if and only if $q_1 = \cdots =q_k$.
Applying it to the denominators of the right hand side of
Eq.~(\ref{eq:hdet_em2}), we find that 
\begin{equation}
{\rm  r.h.s.} \leq 
\frac{\left(\alpha_0 \cdots \alpha_{k_1 -1}\right)^{\frac{2}{k_1}} +
      \left(\beta_0 \cdots \beta_{k_1 -1}\right)^{\frac{2}{k_1}} }
{k_1^{\frac{d-2}{2}} 
\left(|z_0|\cdots |z_{k_1 -1}|\right)^{\frac{d-2}{k_1}}}
\leq 1 ,
\end{equation}
with both the equalities attained for 
\begin{align}
|z_0| &= \cdots = |z_{k_1 -1}| = \frac{1}{\sqrt{k_1}} ,\\
\alpha_0 &= \cdots = \alpha_{k_1 -1}.
\end{align}
This completes the proof.

\section{Implications to 2-qubit mixed states}

Let us briefly discuss the implications to 2-qubit mixed states.
If the third subsystem is assumed as an "environment" for 2-qubit
mixed states, we can intuitively expect that the 2-qubit substructure of 
Alice and Bob is more controllable
in the $2 \times 2 \times n$ pure case, than in the 2-qubit mixed case. 
Indeed, our result implies that the intuition is correct, and we will
see that even when the third "fairy" Clare helps Alice and Bob by 
controlling her environment as they desire, there remains a limitation 
in conversion of the 2-qubit substructure.

The (single copy of) generic 2-qubit mixed states $\rho $ of rank 4 
can be transformed to the Bell diagonal forms under the invertible local 
operations \cite{linden98,verstraete01}.
Let us associate $\rho$ with its purified state $|\Psi\rangle$
in the $2 \times 2 \times n$ format, i.e.,
\begin{equation}
\rho = {\rm tr}_{3}(|\Psi\rangle\langle\Psi|)
=\tilde{\Psi}\tilde{\Psi}^{\dag}.
\end{equation}
Corresponding to three relative weights of the four Bell components
of (normalized) $\rho$ after invertible local operations, 
three relative ratios of the singular values $s_i \,(i=0,\ldots,3)$ of 
\begin{equation}
\label{eq:RTR}
R^T R = \tilde{\Psi}^T (i \sigma_y \otimes i \sigma_y) \tilde{\Psi}
\end{equation} 
in Eq.~(\ref{eq:R}) are the SLOCC invariants of  
2-qubit mixed states (we do not consider the mixing here) \cite{linden98}.
On the other hand, in our $2\times 2\times n$ pure states, 
we are just concerned about whether some of singular values $s_i$
are zero or not, i.e., about the rank of $R^T R$ as a SLOCC invariant, 
in addition to local ranks (cf. the Section~4).

For the degenerate 2-qubit mixed states, their SLOCC conversion is a bit 
complicated.
There are 2-qubit mixed states, often called quasi-distillable states 
\cite{horodecki99,verstraete01}, that can not convert to the Bell diagonal 
forms under invertible local operations, but can approach arbitrarily close 
to them. Still, since we can suitably define the singular values for these 
states, the above argument for generic mixed states holds true.
The exotic property of the quasi-distillable mixed states survives 
in the "limit" to the situation as $2 \times 2 \times n$ pure states. 
An example is 2-qubit mixed states associated with the 
$2 \times 2 \times n$ pure states of the W class.
In turn, even when Alice and Bob are helped by Clare,
they can not convert to the major region  
associated with the GHZ class although they can do arbitrarily close to it.
Finally, we would say, this is why there are more than one classes 
in the middle grades of the partial order of $2 \times 2 \times n$  
multipartite entanglement in Fig.~\ref{fig:part}.

\section{Conclusion}

Mysteries in the fundamental of quantum theory
as well as remarkable potential applications in quantum information 
processing would remain rooted in the rich properties of multipartite 
entanglement.
Then, SLOCC can be utilized as a convenient and useful set of 
quantum operations.
We have presented a brief overview about several characteristics of 
multipartite entanglement, illustrating the 2-qubit and the rest
($2 \times 2 \times n$) quantum system. 
We have shown that its five-graded partially ordered structure of nine 
SLOCC entangled classes causes various multipartite phenomena,
which cover entanglement sharing as fundamental and multi-party LOCC
protocols, for example entanglement swapping, as applications.

%%%%%%%%%%%%%%%%%%%%%%%%%%%%%%%%%%%%%%%%%%%%%%%%%%%%%%%%%%%%
% Doing Acknowledgement                                    %
%%%%%%%%%%%%%%%%%%%%%%%%%%%%%%%%%%%%%%%%%%%%%%%%%%%%%%%%%%%%

\section*{Acknowledgments}

The author would like to thank F.~Verstraete, a talk in the EQIS 
conference (September, 2003, Kyoto, Japan) was based on the joint work with, 
for reading the manuscript and helpful feedback.
He also appreciates enjoyable discussions with the participants of 
the EQIS'03, in particular, V.~Bu\v{z}ek and M.M.~Wolf about the 
shareability of multipartite entanglement.
The work is partially supported by the Grant-in-Aid for JSPS Fellows.

%%%%%%%%%%%%%%%%%%%%%%%%%%%%%%%%%%%%%%%%%%%%%%%%%%%%%%%%%%%%
% Doing Appendix(ices) - Appendix A & B are shown below    %  
%%%%%%%%%%%%%%%%%%%%%%%%%%%%%%%%%%%%%%%%%%%%%%%%%%%%%%%%%%%%

\appendix

\section{Shareability of multipartite entanglement}

The classifications of entanglement are quite relevant and complementary to the
studies of entanglement sharing.
The shareability of entanglement is qualitatively a monogamous nature of 
entanglement \cite{terhal03}. For example, if a 2-qubit system is 
the maximally entangled pure state (a Bell pair) then neither of 2 qubits 
can entangle with the other quantum system.
Since there is no such a limitation for correlations in {\it classical} 
multipartite systems, this nature is the very quantum effect, and is also 
expected to have many potential applications to QIP. 

However, the {\it quantitative} understanding can be
said to be still limited despite a lot of efforts.
Exceptionally, when the whole quantum system is a 3-qubit pure state 
$|\Psi\rangle$, a constraint of the shareability \cite{coffman00} 
is expressed as
\begin{equation}
\label{eq:ent_share}
C_{(12)3}^2 \equiv C_{13}^2 + C_{23}^2 + 4 |{\rm Det}_{222}\Psi|.
\end{equation}
The concurrence $C$ is the measure of entanglement of formation 
for 2-qubit mixed states $\rho$ \cite{wootters97}, and is given by
\begin{equation}
C={\rm max}(
s_0^{\downarrow}-s_1^{\downarrow}-s_2^{\downarrow}-s_3^{\downarrow},\; 0),
\end{equation}
where $s_i^{\downarrow} \;(i=0,\ldots,3)$ are the square root of 
decreasing-ordered eigen values of a non-Hermitian positive operator
$\rho (i\sigma_y \otimes i\sigma_y) \rho^{\ast} (i\sigma_y \otimes i\sigma_y)$
(the superscript $\ast$ stands for the complex conjugate), or equivalently,
the decreasing-ordered singular values of $R^T R$ of Eq.~(\ref{eq:RTR}).

Although Eq.~(\ref{eq:ent_share}) was first considered as the definition of
a {\em residual} entanglement, called the 3-tangle 
$\tau=4|{\rm Det}_{222}\Psi|$, it can be seen as an identity between 
different kinds of entanglement. 
Eq.~(\ref{eq:ent_share}) suggests that when we consider its left hand side
represents the entire entanglement between Clare and the rest,
there is a trade-off between the amount of genuine tripartite entanglement, 
given by $4|{\rm Det}_{222}\Psi|$, and the amount of pairwise entanglement,
given by $C_{13}$ and $C_{23}$. 
The GHZ and W representative states appear as the extremal states of 
the trade-off constraint in a straightforward manner.
The GHZ state $|000\rangle+|111\rangle$ maximizes the genuine tripartite
component 3-tangle ($\tau=4|{\rm Det}_{222}\Psi|$), and accordingly
has no pairwise component $C_{j j'} \;(j,j'=1,2,3)$ at all.
On the other hand, the W state $|001\rangle+|010\rangle+|100\rangle$
has no genuine tripartite component, but maximizes the total amount of
pairwise component $C_{12}^2 + C_{23}^2 +C_{31}^2$ \cite{dur00,koashi00}.
As a nontrivial fact, it is interesting to observe that the intermediate 
entangled states are equivalent to the GHZ state under the SLOCC equivalence.

%%%%%%%%%%%%%%%%%%%%%%%%%%%%%%%%%%%%%%%%%%%%%%%%%%%%%%%%%%%%
% Doing references:                                        %
%%%%%%%%%%%%%%%%%%%%%%%%%%%%%%%%%%%%%%%%%%%%%%%%%%%%%%%%%%%%

\end{document}